*"Quest for the non-perturbative magnetic field effects in 1000-Tesla"*

# Non-linear transport in field-induced insulating states of graphite

Masashi Tokunaga[1]*, Kazuto Akiba[2], Hiroshi Yaguchi[3], Akira Matsuo[1], and Koichi Kindo[1]

[1]*The Institute for Solid State Physics (ISSP), The University of Tokyo, Kashiwa, Chiba 277-8581, Japan.*
[2]*Faculty of Sci. and Eng., Iwate University, Morioka, Iwate 020-8551, Japan.*
[3]*Dept. of Physics and Astronomy, Tokyo Univ. of Sci., Noda, Chiba 278-8510, Japan.*

*E-mail: tokunaga@issp.u-tokyo.ac.jp

**Abstract.**   Graphite exhibits multi-stage phase transitions in the quantum-limit states realized by magnetic fields applied along the *c*-axis. Despite extensive studies on this phenomenon, the origin remains a matter of debate to this day. We performed high-field magnetotransport measurements on single crystals of graphite, focusing on the non-linear conductivity in pulsed-magnetic fields of up to 75 T. The longitudinal magnetoresistance exhibits distinct non-linearity not only in the first but also in the second field-induced phases.

## 1. Introduction

In sufficiently high magnetic fields where cyclotron energy dominates over the Fermi energy, all charge carriers in metals fall into the lowest Landau subband. In such a quantum limit state, the effects of electron correlations become more pronounced in relation to the reduced kinetic energy. Theoretical studies have predicted the emergence of anomalous quantum phases due to the many-body effect [1,2]. On the other hand, experimental realization of such quantum states is a challenging subject because it requires extremely high magnetic fields for ordinary conductors. In this context, semimetallic graphite has garnered significant attention as a suitable platform for this subject.

Semimetallic graphite with a layered structure enters the quasi-quantum limit state, where spin-split lowest electron and hole Landau subbands only intersect the Fermi level, in a magnetic field greater than 7 T applied along the out-of-plane (*c*-axis) direction. A further increase in the magnetic field causes a steep increase in resistance at low temperatures in fields approximately higher than 30 T [3]. Since the critical field exhibits significant temperature dependence, this steep increase in resistance is ascribed to a phase transition resulting from a many-body effect.

Graphite in the quasi-quantum limit state is a quasi-one-dimensional conductor that contains eight Fermi points in the first Brillouin zone. Accordingly, the system is unstable against nesting with the wave number vectors connecting these points. Depending on the choice of the nesting vector, the stabilized state can be classified into charge density waves, spin density waves, or excitonic phases. Recent experiments, extending up to a field of 90 T, have revealed the emergence of multiple phase transitions, suggesting the realization of distinct phases in different magnetic field regions, while the origin remains unidentified [4-12].

Experimental techniques capable of elucidating the physical origins of the field-induced phases are limited at high magnetic fields, making measurements of non-linear conduction potentially valuable for providing unique insights into these phenomena. Iye *et al*. revealed non-linearity in the in-plane conductivity within the first field-induced phase (phase A) [13]. Later, Yaguchi *et al*. elucidated that more pronounced non-linearity emerges in the out-of-plane conduction [14]. These experimental results are reminiscent of the sliding phenomenon associated with a charge density wave state, characterized by charge modulation along the *c*-axis direction.

On the other hand, Zhu *et al*. recently proposed a new interpretation suggesting that the A phase originates from an excitonic phase [10]. Furthermore, through magnetoresistance measurements in magnetic fields of up to 90 T, they claimed the existence of other field-induced phases in the field range from 53 T to 75 T (phase B) and beyond 75 T (phase C) [11]. The physical origins of each phase remain unclear, necessitating further experiments to elucidate their underlying mechanisms. Therefore, we studied non-linear transport properties of graphite in magnetic fields of up to 75 T.

## 2. Experimental methods

We studied the magneto-transport properties of single crystals of Kish graphite under magnetic fields applied along the *c*-axis of the crystals. Electric contacts for out-of-plane resistance ($R_c$) measurements were formed as described in Ref. 14. Pulsed magnetic fields up to 56 T and 75 T were generated using non-destructive pulse magnets with durations of 36 ms and 4 ms, respectively, installed at the International MegaGauss Science Laboratory of ISSP [15].

## 3. Results and discussion

Figure 1 shows out-of-plane magnetoresistance measured up to 56 T. At 4.2 K, $R_c$ steeply increases at 34 T and then decreases at 53 T, which represents the emergence of phase A in this intermediate field region. The enhancement of $R_c$ in phase A becomes more prominent with decreasing temperature. At 1.5 K, we observed positive magnetoresistance above 53 T, which corresponds to the emergence of phase B in this field region.

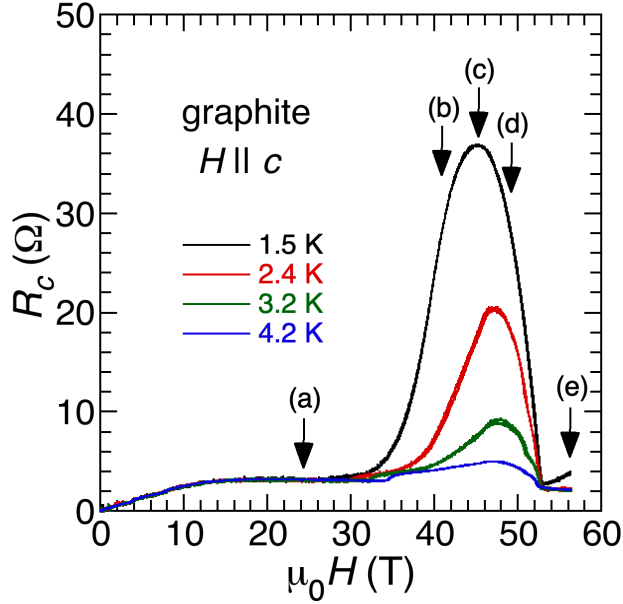

**Fig. 1.** Out-of-plane magnetoresistance of graphite up to 56 T measured at several temperatures between 1.5 K and 4.2 K. The arrows indicate the fields where non-linearity was examined.

Figures 2(a)-(e) show the relation between current ($I$) and electric field ($E$) at several temperatures. At a field of 24.0 T, the $I$-$E$ profiles exhibit an almost linear relation at all the temperatures between 1.6 K and 4.2 K. As the field increases to 41.2 T, superlinear enhancement of $I$ is observed for larger $E$ at 1.3 K, as shown in Fig. 2(b). Such non-ohmic behavior is also recognized in phase A at fields of 45.5 T and 50.0 T [Figs. 2(c) and (d)]. In addition, the $I$-$E$ curve at 56.3 T (in phase B) also exhibits a similar superlinear trace at the lowest temperature [Fig. 2(e)].

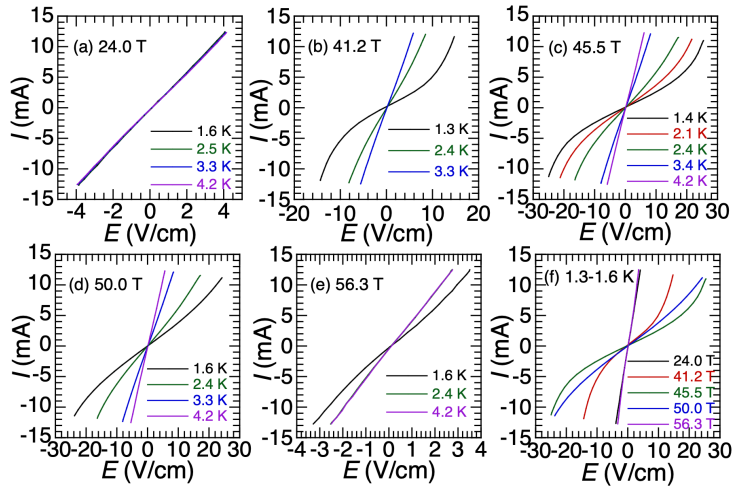

**Fig. 2.** $I$-$E$ curves measured at the top of pulsed magnetic fields with the maximum fields of (a) 24.0 T, (b) 41.2 T, (c) 45.5 T, (d) 50.0 T, and (e) 56.3 T. (f) $I$-$E$ curves measured at temperatures between 1.3 K and 1.6 K at various fields.

Non-linearity in the out-of-plane conduction was further studied up to 75 T. Since the field duration is limited in this short-pulse magnet, we could not obtain reliable *I-E* curves at the top of pulsed fields. Alternatively, we measured magnetoresistance with different applied currents. As shown in Fig. 3, at a current of 0.2 mA, the out-of-plane resistance shows a double peak structure with the peak fields of 47 T and 68 T corresponding to the field-induced phases A and B as reported earlier (thin dashed lines in Fig. 3). On the other hand, $R_c$ measured at larger current of 0.5 mA shows reduced peak structures both in phases A and B, which represents non-linear behavior in both phases.

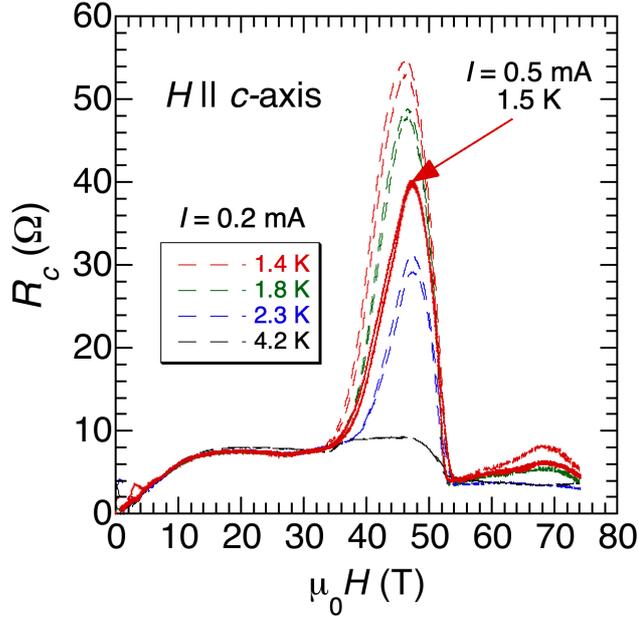

**Fig. 3.** Out-of-plane magnetoresistance of graphite up to 75 T. Thin dashed-lines represent the results obtained for *I* = 0.2 mA, whereas thick solid line shows the data for *I* = 0.5 mA.

Let us consider the origin of non-linear conduction observed in the phase A. Enhancement of the current at higher electric fields is consistent with the sliding phenomenon in the density wave state. According to previous studies, the threshold electric field of non-linearity ($E_0$) depends on magnetic field and temperature [13,14]. Especially, $E_0$ increases linearly as $T/T_c(H)$ decreases from 1, where $T_c(H)$ denotes the transition temperature at a magnetic field of *H*. Several experiments have clarified that $T_c(H)$ reaches its highest value at 47 T and decreases to 0 at 53 T [5,10-12]. Therefore, the sliding scenario predicts a reduction of $E_0$ as the field approaches 53 T. Our experimental data at the lowest temperature [Fig. 2(f)] shows that $E_0$ increases with increasing magnetic field from 41.2 T to 45.5 T, whereas it does not detect a reduction of $E_0$ at a field of 50.0 T contrary to the expectation in the sliding phenomenon of the density wave state.

It is unclear whether non-linearity emerges in the excitonic phase or not. Naively, we do not expect collective motion of charge carriers in this state and anticipate the absence of the non-linearity in the excitonic phase, as previously claimed [13]. Recently, Nakano *et al.* studied a candidate material for the excitonic phase, $Ta_2NiSe_5$, and concluded that the conduction remains ohmic up to 200 V/cm [16]. If non-linear conduction emerges in the excitonic phase, it will not be caused by the collective motion of charge carrier but by the breaking of charge-neutral pairs. If this is the case, the threshold electric field should monotonically increase as the field exceeds 47 T, which is consistent with the results shown in Fig. 2(f). The current dataset is insufficient to establish a conclusive interpretation. Further systematic studies of such non-linearity will be needed at extended field ranges to elucidate the origin of multi-stage phase transitions in graphite.

## 4. Conclusions

We studied the longitudinal magnetoresistance of Kish graphite in magnetic fields of up to 75 T applied along the *c*-axis of the crystals. We observed non-linear transport properties in the first field-induced phase between 30 T and 53 T, as reported previously, and also in the second phase above 53 T. Although the enhancement of the conductivity at elevated electric fields is consistent with the sliding phenomena in density wave states, further studies are needed to elucidate the physical origin of the field-induced phases in graphite.

## Acknowledgments
This work was supported by JSPS KAKENHI Grant No. JP23H04862.

## References

[1] H. Fukuyama, Solid State Commun. **26**, 783 (1978).
[2] B. I. Halperin, Jpn. J. Appl. Phys. **26**, 1913 (1987).
[3] S. Tanuma, R. Inada, A. Furukawa, O. Takahashi, Y. Iye, and Y. Onuki, *Physics in High Magnetic Fields* (Springer, 1981), p. 316.
[4] D. Yoshioka and H. Fukuyama, J. Phys. Soc. Jpn. **50**, 725 (1981).
[5] H. Yaguchi and J. Singleton, Phys. Rev. Lett. **81**, 5193 (1998).
[6] H. Yaguchi and J. Singleton, J. Phys. Condens. Matter **21**, 344207 (2009).
[7] B. Fauqué, D. LeBoeuf, B. Vignolle, M. Nardone, C. Proust, and K. Behnia, Phys. Rev. Lett. **110**, 266601 (2013).
[8] K. Akiba, A. Miyake, H. Yaguchi, A. Matsuo, K. Kindo, and M. Tokunaga, J. Phys. Soc. Jpn. **84**, 054709 (2015).
[9] F. Arnold, A. Isidori, E. Kampert, B. Yager, M. Eschrig, and J. Saunders, Phys. Rev. Lett. **119**, 136601 (2017).
[10] Z. Zhu, R. D. McDonald, A. Shekhter, K. A. Modic, F. F. Balakirev, and N. Harrison, Sci. Rep. **7**, 1733 (2017).
[11] Z. Zhu, P. Nie, B. Fauqué, B. Vignolle, C. Proust, R. D. McDonald, N. Harrison, and K. Behnia, Phys. Rev. X **9**, 011058 (2019).
[12] J. Wang, P. Nie, X. Li, H. Zuo, B. Fauqué, Z. Zhu, and K. Behnia, PNAS **117**, 30215 (2020).
[13] Y. Iye and G. Dresselhaus, Phys. Rev. Lett. **54**, 1182 (1985).
[14] H. Yaguchi, T. Takamasu, Y. Iye, and N. Miura, J. Phys. Soc. Jpn. **68**, 181 (1999).
[15] A. Miyata, K. Matsui, A. Matsuo, A. Kikuchi, and K. Kindo, IEEE Trans. Appl. Super. **36**, 4300204 (2026).
[16] A. Nakano, U. Maruoka, H. Taniguchi, and I. Terasaki, J. Phys. Soc. Jpn. **89**, 045001 (2020).